\documentstyle[12pt]{article}

\newcommand{\eqr}[1]{(\ref{#1})}
\newcommand{\dbydx}{{d\over d x}}
\newfont{\feff}{cmti10}

\def\undertext#1{\vtop{\hbox{#1}\kern 1pt \hrule}}
\def\ra{\rightarrow}
\def\eff{\hbox{\feff eff}}
\def\bra#1{\left\langle #1\right|}
\def\ket#1{\left| #1\right\rangle}

\def\psiZero{\psi_0}
\def\PsiZero{\Psi_0}
\def\braZero{\bra{\psi_0}}
\def\ketZero{\ket{\psi_0}}
\def\VEV#1{\left\langle #1\right\rangle}
\def\tr{\hbox{tr}\,}
\def\be{\begin{equation}}
\def\ee{\end{equation}}
\def\bea{\begin{eqnarray}}
\def\eea{\end{eqnarray}}
\def\eqref#1{\ref{(#1)}}

\def\ddlsi{\frac{\partial^2}{\partial\lambda_i^2}}
\def\ddys{\frac{\partial^2}{\partial y^2}}
\begin{document}
\begin{flushright}PUPT-1354\\ TAUP-2013-92\\
hep-th/9212114\\
December 1992
\end{flushright}
\vskip1cm
\begin{center}
{\LARGE GROUND STATE OF 2D QUANTUM}
\vskip.2cm
{\LARGE GRAVITY AND SPECTRAL DENSITY}
\vskip.2cm
{\LARGE OF RANDOM MATRICES}
\vskip1cm
{\large Marek Karliner $^{\ast}$}
\vskip0.3cm
{\large Alexander Migdal $^{\dagger}$}
\vskip0.3cm
{\large Boris Rusakov $^{\ast}$}
\vskip0.7cm
{\it $^{\ast}$ School of Physics and Astronomy
\\ Raymond and Beverly Sackler Faculty of Exact Sciences
\\ Tel-Aviv University, 69978 Tel-Aviv, Israel }
\vskip0.5cm
{\it $^{\dagger}$ Physics Department and
\\ Program in Applied and Computational Mathematics
\\ Fine Hall, Princeton University
\\ Princeton, NJ 08544-1000}
\end{center}
\vskip 1.5cm
\begin{abstract}
We compute the exact spectral density of random matrices in the
ground
state of the quantum hamiltonian corresponding to the matrix model
whose double scaling limit describes pure gravity in 2D.  We show
that
the non-perturbative effects are very large and in certain cases
dominate the semi-classical WKB contribution studied in the earlier
literature.  The physical observables in this model are the loop
averages with respect to the spectral density. We compute their exact
ground-state expectation values and show that they differ
significantly from the values obtained in the WKB approximation.
Unlike the alternative regularizations of the nonperturbative 2D
quantum gravity, based on analytic continuation of the Painlev\'e
transcendent, our solution shows no pathologies.
\end{abstract}

\newpage

The discovery \cite{DSL} of the double scaling
limit of the matrix models \cite{MatMod} of 2D Gravity and its
relation \cite{KdV} to the KdV hierarchy opened the way to
the study of the nonperturbative phenomena in Gravity and
String theories. These phenomena are supposed to be described
by the so called string equation
\be
x = \sum t_{l} (2 l + 1) R_{l} \left[C \right]
\label{stringeq}
\ee
where $x$ is the scaling combination of cosmological constant
and parameter $ 1/N $ of the genus expansion,
$t_{l}$ are the mass parameters of the theory
and $R_{l}\left[C \right]= C^{l} + {\it derivatives}$ are
the Gelfand-Dikii differential functionals corresponding to the
specific heat $C(x)$ as the potential of the Schr\"odinger
operator ( see e.g. \cite{KdV} for details).

The string equation is in general a differential equation
for $C(x)$ of the order corresponding to the highest $l$ in
the sum. As observed in \cite{BMP}, for odd $l$
there is the natural boundary condition

\be
C(x) \ra x^{1/l}
\qquad\quad x \ra \pm \infty
\label{BC}
\ee
which uniquely determines the solution.

For the even $l$ case situation is more complicated. The above
boundary condition would lead to a complex solution,
corresponding analytic continuation of the divergent matrix
integral \cite{FDavid}. The attempt to define the even $l$
case as the limit $t_{l+1} \ra 0 $ of the odd case
\cite{DSS} does not work, since the limit does not
exist.

One way to interpret this paradox is to claim the even case
and in particular the $l=2$ case of pure 2D~Gravity
inconsistent. That would be an exciting possibility : the
nonperturbative creation of matter in Gravity. However, the
matter here is not quite physical, as it does not correspond
to the unitary multiplets of conformal field theory.
The partition functions of the $l>2$ models with higher genus
are not positive \cite{GM}.

So, in fact, this interpretation implies that no 2D~Gravity
could exist at nonperturbative level. This is too serious a
statement to make without investigating the alternative
interpretations.

One such interpretation, suggested by Marinari and Parisi
\cite{MP}, along the lines of the general method by
Greensite and Halpern \cite{GH}, seems particularly
appealing. They do not add any ad hoc terms to stabilize the
equation, but rather modify the basic definition of the
matrix model to make it meaningful for arbitrary potential.
In Ref.~\cite{MP} the spectral density of the resulting
quantum hamiltonian was studied in the semi-classical WKB
approximation.
It turns out to be possible to go beyond the WKB approximation
and to obtain the exact spectral density of this hamiltonian.
The necessary framework was set up in Ref.~\cite{KarMig},
in which the set of equations
generalizing the string equation was derived and
studied numerically. Quantum effects, invisible in the WKB
approximation, turned out to be extremely important.

The prescription of Refs.~\cite{MP},\cite{GH} has attracted
a considerable amount of attention, \cite{AGV}-\cite{Cosmo}.
In particular, there have been some interesting attempts
to study the properties of the model numerically
at finite $N$. The problem encountered
in these studies is the same as with all numerical studies of
the matrix models, namely the double scaling limit requires
very large $N$, because of the ${N^{}}^{1/5}$ dependence of
physical observables.

The goal of the present paper is to apply the analytic and numerical
methods of the note \cite{KarMig} to derive the exact
spectral density of random matrices in the ground state of
2D~Gravity.
As a result, we are able to demonstrate that the non-perturbative
effects are very large and in certain cases completely dominate
the semi-classical WKB contribution. The physical observables in
this model are the loop averages with respect to the spectral
density.
 We compute their exact ground-state
expectation values and show that they differ significantly
from the values obtained in the WKB approximation.

We begin by discussing the method of Greensite and Halpern for the
simplest example of a one dimensional integral.
Consider a positive
action $S(x)$ depending on a single variable $x$:
\be
Z=\int d x \exp\left[-S(x) \right]
\label{xAction}
\ee
By definition, the average of an operator $Q(x)$ in the action $S$
is given by
\be \VEV{Q}_Z\equiv{\displaystyle\int d x \,Q(x) \exp[- S(x) ]
\over \displaystyle\int d x \exp\left[-S(x) \right] }
\label{QvevDef}
\ee
If we define
\be
\psiZero\equiv \exp\left[-S(x)/2 \right]/\sqrt{Z}
\label{PsiZeroDef}
\ee
the expectation value \eqr{QvevDef} can be written
\be
\VEV{Q}_Z = \braZero Q \ketZero
\label{QVEV}
\ee
$\psiZero$ is always positive and therefore it is always
possible \cite{Turbiner},\cite{GH},\cite{MP}
to construct a quantum-mechanical hamiltonian
$ H  = -d^2/dx^2 + V(x) $,
such that
$\psiZero$ is its ground-state wave-function with the ground-state
energy equal to zero:
\bea
H\psiZero = 0 \nonumber\\
-\psi_0^{''}(x)+ V(x) \psi_0(x)=0 \\
\label{HofPsi}
V(x)= \psi_0^{''}(x)/\psi_0(x)\nonumber
\eea
It is instructive to rewrite this hamiltonian in
a manifestly positive form, as
a product of an operator and its hermitean conjugate,
\be
H=\left(-\dbydx + {S^{'}(x)\over 2} \right)
  \left( \dbydx + {S^{'}(x)\over 2} \right)
\label{Hdef}
\ee
$\psiZero$ is annihilated by the second factor and hence
$H\psiZero=0$. When the action
$S$ is unbounded from below, the average \eqr{QvevDef}
is ill-defined. Formally, \eqr{PsiZeroDef} is still annihilated by
\eqr{Hdef}, but $\psiZero$ is now not normalizable and therefore
cannot be the ground state of $H$. On the other hand,
$H$ is a positive hamiltonian by construction
and therefore a normalizable ground state $\PsiZero$
must exist. Clearly the true ground state $\PsiZero$
must be different from
$\psiZero$ and must have an eigenvalue $e_0 >0$.
Consider the action $S(x)=x^2 - g^2 x^4$. The average
\eqr{QvevDef} is ill defined,
but the power-series expansion in $g^2$ exists and can
be computed explicitly.
Each order in the
expansion in powers of $g^2$ involves only moments of $x$
with a gaussian measure,
\bea
\exp[-S(x)]=\exp(-x^2) \sum_{k=0}^{\infty}{(g^2 x^4)^n\over n!}
\nonumber\\
\VEV{Q}^l_Z =
{\displaystyle\int d x \,Q(x) \exp(-x^2)\, \sum_{n=0}^l{(g^2 x^4)^n/
n!}
\over \displaystyle\int d x   \exp(-x^2)\, \sum_{n=0}^l{(g^2 x^4)^n/
n!}}
\label{VEVQn}
\eea
Any sensible definition of average with the action $S(x)$
ought to reproduce the perturbation expansion \eqr{VEVQn}
and it should reduce to \eqr{QvevDef} for a positive action.
The quantum-mechanical expectation value in the {\em true}
ground-state $\PsiZero$ of $H$ satisfies both constraints,
we therefore take it as the {\em definition} of the average with the
action $S$:
\be
\VEV{Q}_Z = \bra{\PsiZero} Q \ket{\PsiZero}
\label{QVEVtrue}
\ee
This prescription can be applied to matrix models,
\be
e^{F}=Z= \int d \phi \exp\left[  -\beta \tr U_k(\phi) \right]
\label{ZmatrixDef}
\ee
where $\phi$ is a hermitean $N\times N$ matrix and the critical
potentials are given by
\be
U_k(\phi) = \int_0^1 \frac{d t}{t}
\left[ 1 - \left( 1 - t (1-t) \phi^2 \right)^k \right]
\label{Ukphi}
\ee
At large $\phi$ the $\phi^{2k}$ term dominates, with the
coefficient $(-1)^{k+1} \int_0^1 d{t}\,t^{k-1} (1-t)^k$. Therefore
the $k$-even critical potentials are unbounded from below and
it is necessary to provide a prescription for defining the
average \eqr{ZmatrixDef} beyond perturbation expansion.
Following the work of Greensite and Halpern \cite{GH} and
Marinari and Parisi \cite{MP},
we adopt the prescription \eqr{QVEVtrue}.
After all, the only existing justification of the matrix models
as theories of gravity is perturbative, in the sense of the genus
expansion.
The quantum mechanical definition is \hbox{\em a priori} as
good as the statistical one,
but has the advantage of being guaranteed to make sense for all
models.
The extra bonus of this prescription is the relation to the one
dimensional supersymmetric string theory \cite{MP}, with
dynamically broken supersymmetry
in the even $l$ case. For practical purposes the
supersymmetry seems to be useless so far, but one may hope to relate
it to
the physical supersymmetric string theories.

The simplest case of an unbounded potential in a
matrix model is the cubic potential
\be
U(\phi)= \frac{\phi^2}{2} - g \frac{\phi^3}{3}
\label{Ucubic}
\ee
In order to proceed, it is useful to change variables from
the hermitean $N\times N$ matrices to their eigenvalues,
$\left\{\lambda_1,\dots,\lambda_N\right\}$:
\be
e^{F}=Z= \int \prod_i d \lambda_i
\prod_{i<j} (\lambda_i -\lambda_j)^2
 \exp\left[  -\beta \tr U(\lambda)\, \right] \nonumber\\
\label{ZmatrixReDef}
\ee
The factor
$\displaystyle{\prod_{i<j}} (\lambda_i -\lambda_j)^2$
comes from the Jacobian.
In the following it is useful to include this factor in the
wavefunction. The integration measure is then simply
$\displaystyle{\int \prod_i} d \lambda_i$.
It follows that the effective wave function corresponding to
\eqr{ZmatrixReDef} is
\be
\Psi^{\eff} = \prod_{i<j} \left(\lambda_i - \lambda_j\right)
\psi(\lambda)/\sqrt{Z}
\label{Psieff}
\ee
The hamiltonian $H$ is immediately obtained from \eqr{Hdef} and
\eqr{ZmatrixReDef}:

\be
H=\sum_i H_i\,;\qquad
H_i =-\ddlsi + V_{\eff}(\lambda_i)\,;\qquad
V_{\eff}(\lambda)=N \beta \left( g \lambda - \frac{1}{2} \right)
+ \frac{\beta^2}{4}\left( \lambda- g \lambda^2\right)^2
\label{hFermi}
\ee
\eqr{Psieff} and \eqr{hFermi} describe
the \undertext{ideal}  Fermi gas of $N$ non-interacting
particles. The fermionic property is generic -- it is due to the
Jacobian of the transformation from the matrices $\phi$ to their
eigenvalues. On the other hand,
the fact that the effective
fermions are free is only true
when the original potential $U(\phi)$ is cubic in $\phi$.

The effective potential depends on the free variable $\lambda$
and on three parameters of the original matrix-model action:
$N$, $\beta$ and $g$,
$V_{\eff}=V_{\eff}(N,\beta,g;\,\lambda)$.
We are interested in the critical properties
of the theory in the double scaling limit,
\be
x={\beta-N\over N^{1/5}}\,; \qquad N\ra\infty\,;\qquad x\sim N^0
\label{ScalingLimit}
\ee
with the critical point at $x=0$.
For small positive $x$ the effective potential
$V_{\eff}(N,\beta,g;\,\lambda)$ has a double-well shape.
In the double scaling limit \eqr{ScalingLimit} the depth of
the left well scales like $\sim N^2$, while the depth of the
right well scales like $\sim x^{3/2} \sim N^0$. Despite this,
it is the right, tiny well, which is responsible for all the
interesting critical phenomena in the double scaling limit.
All the cubic potentials of the type \eqr{Ucubic} are in the
same universality class. At the critical point the depth of the
second well goes to zero, and the potential has an inflection point,
instead of a double minimum.

Since we are interested in having the critical point at $x=0$,
it is necessary
to chose a value of $g$ such that for $x=0$, i.e. for $\beta=N$,
the extremum is an inflection point:
\be
\left.
{\partial
V_{\eff}\over\partial\lambda}\right\vert_{\lambda=\lambda_0}
=0 \,;\qquad
\left.
{\partial^2 V_{\eff}\over\partial\lambda^2}
\right\vert_{\lambda=\lambda_0}
 = 0
\label{ExtremumCond}
\ee
leading to
\be
g={1\over\sqrt{12\sqrt{3}}}
\label{gcritical}
\ee
\be
\lambda_0 = \left( \sqrt{3} + 1\right) \root 4 \of 3
\ee
We are interested in the scaling properties of the Fermi energy,
or ground state,
of the hamiltonian \eqr{hFermi}. The scaling
properties are determined by the tiny right well, but the bulk
of energy levels is in the left, bigger well. The Fermi
energy, $e_F$ can be written as
\be
e_F = e_F^0 + e_F^s(x)
\label{eFermiI}
\ee
where the  bulk part, $e_F^0\ $, is
determined by the left well and
$e_F^s(x)$ is the $x$-dependent scaling part
that we are interested in. The depth of the left well
$\sim N^2$, while the depth of the right well $\sim N^0$.
In terms of magnitude,
$e_F$ is completely dominated by $e_F^0$.
The hamiltonian \eqr{hFermi} can only be solved by some
approximation procedure, analytical or numerical.
It is mandatory to first isolate and ``magnify" the scaling
part, otherwise any approximate result for $e_F$
will be dominated by $e_F^0$ and the $x$ dependence of the
fine-structure will be lost.

To isolate the scaling part, we expand the effective potential
in $x$ and in $\lambda-\lambda_0$
around the inflection point and obtain
the scaling potential $v(y)$:
\be
V_{\eff}(N,\beta,g;\,\lambda)=
V_{\eff}(N{=}\beta,g;\,\lambda{=}\lambda_0)
+ \beta^{4/5}\alpha^{2/5}v(y)
\label{VeffZero}
\ee
\be
v(y)={y^3\over3}-\epsilon\, y
\label{vofy}
\ee
where $x$ is given by \eqr{ScalingLimit} and
\be
y=(\lambda-\lambda_0)\alpha^{1/5}\beta^{2/5},\qquad
\epsilon=x\alpha^{-3/5},\qquad
\alpha={2+\sqrt{3}\over 4 \root 4 \of 3}
\label{NewVars}
\ee
The critical properties of the theory are thus
determined by the scaling hamiltonian
\be
h= -\ddys + v(y)
\label{scalingH}
\ee
The cubic potential $v(y)$ (not to be confused with the original
cubic potential \eqr{Ucubic} ) is formally
unbounded from below. This is not a problem, however, since $v(y)$
is only meaningful in the scaling region  and the full
potential $V_{\eff}(\lambda)$ is bounded.
The strategy for computing the fine structure of the Fermi
energy is then as follows.
An exact numerical
solution for the spectral density of the scaling hamiltonian
\eqr{scalingH} can be obtained through the powerful methods of
Gelfand and Dikii \cite{GeD}.
The exact solution can then be compared with the spectral
density obtained from the WKB approximation. The two differ
 in the scaling region only, since WKB
is completely adequate in the large left well.
Their difference converges fast,
yielding the contribution of the scaling region.

In order to obtain the density of states,
we first write down the general equations of the
Fermi-gas theory in one dimension.
It is convenient to scale out $ \beta $
from the Hamiltonian, and introduce the resolvent,
\be
G(e,y)= \bra{y}{(h-e)}^{-1}\ket{y}
\label{resolvent}
\ee
The particle density $ \rho(e,y) $ is related to
the imaginary part of the  resolvent,
\be
\rho(e,y)={1\over \pi}\, \hbox{Im}\, G(e+i0,y)
\label{rhoImG}
\ee
and the spectral density  $\nu(e)$ is given by the integral of $\rho$
\be
\nu(e) = \int_{-\infty}^{\infty} d y \,\rho(e,y).
\label{nurho}
\ee
The normalization of $\nu(e)$ is fixed by the fact that
\eqr{Psieff} and \eqr{hFermi} describe
$N$ non-interacting fermions.
We rescale the spectral density by $N$ and
the equation for the Fermi energy $e_F$ is therefore
\be
1 = \int_{-\infty}^{e_F} d e\, \nu(e).
\label{EFeq}
\ee
The WKB particle density is given by
\be
\rho_{WKB}(e,y)=
{1 \over {2 \pi \sqrt{e-v(y)}}}
\label{rhoWKB}
\ee
with the corresponding WKB spectral density
$\nu_{WKB}$
\be
\nu_{WKB}(e) = \int_{-\infty}^{y_1} d y \,
{1 \over {2 \pi \sqrt{e-v(y)}}}
\label{nuWKB}
\ee
where $y_1$ is the first root of $e-v(y)$.
$\nu_{WKB}(e)$ can be expressed in terms of elliptic integrals.
When there are 3 real roots $y_1<y_2<y_3$ of $v(y)-e=0$,
from eqs.~(3.131.1) and (8.112.1) of Ref.~\cite{GR} we obtain
\be
\nu_{WKB}(e) = {1 \over \pi \sqrt{y_3-y_1}} \,{\bf K}(p);\qquad
p=\sqrt{{y_3-y_2\over y_3-y_1}}
\label{nuWKBthree}
\ee
When there is one real root $y_1$ and two complex-conjugate
roots $y_2^*=y_3$,
one can use eq.~(8.126.1) of Ref.~\cite{GR} to transform
\eqr{nuWKBthree}
into the form
\be
\nu_{WKB}(e) = {1 \over \pi \sqrt{\eta} }\,
{\bf K}(\sin(\phi/2));\qquad
y_2-y_1=\eta e^{i \phi}
\label{nuWKBone}
\ee
The ``Fermi energy"
corresponding to the WKB solution, $e_F^{WKB}$ is defined by
\be
1 = \int_{-\infty}^{e_F^{WKB}} d e\, \nu_{WKB}(e).
\label{efWKB}
\ee
Outside the scaling region the exact spectral
density is equal to the WKB density,
\be
\nu(e)=\nu_{WKB}(e)\qquad \hbox{for}\quad e<e_{low},
\label{asymptotics}
\ee
 where $e_{low}$ is some
large negative value of $e$. The WKB solution for the Fermi energy
 $e_F^{WKB}$  for $ \epsilon >0$ was found in
\cite{MP}. It exactly coincides with the bottom of the second
well,
which in our normalization is
\be
 e_F^{WKB} = -{2\over3}\epsilon^{3/2} .
\label{efWKBvalue}
\ee
Combining \eqr{EFeq}, \eqr{efWKB},
\eqr{asymptotics} and \eqr{efWKBvalue}, we obtain the final implicit
equation for the Fermi energy $e_F$ as determined by
the scaling region:
\be
\int_{e_{low}}^{e_F^{WKB}} d e\, \nu_{WKB}(e) =
\int_{e_{low}}^{e_F} d e\, \nu(e) .
\label{ImplicitEf}
\ee
The exact solution for spectral density can be obtained from
the Gelfand-Dikii equation \cite{GeD} for the resolvent $G(e,y)$
\be
- 2 G \partial^2 G/\partial y^2 + {(\partial G/\partial y)}^2 +
4(v-e)G^2 = 1
\label{GeDeq}
\ee
Under usual circumstances, when the parameters are not fine-tuned
to magnify the  scaling region,
the G-D equation is rather useless, as the direct solution of
the Schr\"odinger  equation is simpler.
However, in our case it is just what we need.
Differentiating \eqr{GeDeq} with respect to $y$, dividing by $G$ and
taking the imaginary part, we obtain an ordinary third-order linear
differential equation for the continuous
particle density  $\rho(e,y)$ in the double scaling limit,
\be
\rho^{\prime\prime\prime} = 2 v^{\prime} \rho + 4 (v-e)\rho^{\prime}
\label{rhoeq}
\ee
At large $|y|$,
the general asymptotic form of the solution for $\rho$,
correct up to terms $\sim\,{\cal O}(1/|v|)$,
 can be written
in terms of three integration constants
$c^{\pm}_1$, $c^{\pm}_2$ and $c^{\pm}_3$:
\be
\rho\,\, \ra \,\,
{  c^{\pm}_1 \exp\left( 2 \displaystyle\int^y \sqrt{v} d y \right)
 + c^{\pm}_2 \exp\left(-2 \displaystyle\int^y \sqrt{v} d y \right)
 + c^{\pm}_3
\over \sqrt{\vert v \vert}\,{} }\,;
\qquad\quad y \ra \pm \infty
\label{rhoasymp}
\ee
To chose the proper boundary conditions, physical intuition
about the system must be used.
At $y \to \infty $ the potential $v(y)$ grows large,
so only the decaying exponential is left, while $c^{+}_1=c^{+}_3=0$.
This leaves  one free parameter, $c^{+}_2$,
the overall normalization of $\rho$, which
can be determined as follows.
We start from the asymptotic solution \eqr{rhoasymp}
at some large positive $y=y_0$,
with $c^{+}_1=c^{+}_3=0$ and some
arbitrary initial value of $c^{+}_2$.
We then solve the differential equation \eqr{rhoeq} numerically,
evolving
down to large negative values of $y$, where $v(y) < 0$
and $\sqrt{v}\ $ is complex.
The solution there is of the form \eqr{rhoasymp}, with
$c^{-}_1$, $c^{-}_2$, $c^{-}_3\neq 0$, i.e. it
contains two oscillating exponentials, plus a powerlike term.
The oscillations represent a
pure quantum effect, invisible in WKB expansion. If we average
$\rho \sqrt{(e-v)}$ over these oscillations, the $c^-_1$ and
$c^-_2$ terms disappear and we should obtain
${\displaystyle 1\over\displaystyle {2^{} \pi}}$,
according to the WKB solution.
This means that the solution ought to be multiplied by a constant
such that $c^-_3=1/2\pi$.
This is the  missing normalization condition for the density.

In practice averaging over the oscillations is rather tricky,
as it involves delicate
cancellations between the positive and the negative contributions.
There is, however, a better way of extracting $c^-_3$ from $\rho$.
If $\rho$ is given by \eqr{rhoasymp}, then,
up to terms $\sim\,{\cal O}(1/|v|)$,
\be
c^-_3= \sqrt{e-v}
\left( \rho - {\rho^{\prime\prime}\over4(e-v)}\right)
\label{CIII}
\ee
The prescription \eqr{CIII} has the advantage that it is local in $y$
and only requires the knowledge of
$\rho^{\prime\prime}$, which is readily available in any
code used for solving differential equations.
The reliability of \eqr{CIII} can easily be tested, by verifying
that the result for $c^-_3$ is independent of $y$.

We employed the differential equation solving routine
ODE, described in detail in Ref.~\cite{ODEsolver}.
The routine is very stable
and extremely easy to use.  It is based on a variable-step,
variable-order Adams method (explicit linear multistep method).
The variable-step feature is essential, because of the rapid
crossover
from smooth to oscillatory behavior.

Once  $\rho(e,y)$ is known, $\nu(e)$ is in principle
given by \eqr{nurho}. In practice, it is more efficient to
solve for the integral of $\rho(e,y)$:
\be
\xi(e,y)=-\int_y^{\infty} d y \,\rho(e,y)\,;\qquad\quad
\rho(e,y)=\partial \xi(e,y)/\partial y
\label{xiDef}
\ee
which satisfies the 4-th order differential equation
\be
\xi^{\prime\prime\prime\prime} = 2 v^{\prime} \xi^{\prime} +
4 (v-e)\xi^{\prime\prime}
\label{xieq}
\ee
yielding both $\rho(e,y)$ and $\nu(e)$ at the same time,
\be
 \nu(e)=-\lim_{y\,\ra\, -\infty} \xi(e,y)
\label{nuFromxi}
\ee
The boundary conditions for the derivatives of $\xi$
are those for $\rho$ and its derivatives. In addition,
$\xi(y_0,e)=0$, where  $y_0$ is large and positive.

While solving eq.~\eqr{xieq} for the various values
of $e$, we store
the values of $\rho(e,y)$ on a two-dimensional grid in the
$[e,y]$ plane. These are to  be used later in obtaining
the distribution
of the eigenvalues of a random matrix (see eq.~\eqr{EVdist} below).

The tail of the integral in \eqr{xiDef} converges rather slowly,
since
$\rho(e,y)\sim |y|^{-3/2}$ for large negative $y$.
In that region, however,
the integral of $\rho(e,y)$ is very
well approximated by the integral of $\rho_{WKB}(e,y)$,
eq.~\eqr{rhoWKB}. The corresponding WKB tail $\tau(y)$ can
expressed in terms of elliptic integrals, which we calculated
numerically using standard methods.

The  solution of the G-D equation for
$\rho$ at $\epsilon=1$ and
$ e = -{2\over3}$, which corresponds to the
bottom of the second well is presented in Fig.~1.
The quantum  effects are huge!

Once $\nu(e)$ is known, we obtain the Fermi energy $e_{F}(\epsilon)$
from eq.~\eqr{ImplicitEf}. The result is shown in Fig.~2.

We can now obtain the distribution $\Omega_{\epsilon}(y)$
of the eigenvalues of a random matrix
 in the double scaling limit.
This is done by integrating $\rho(e,y)$ with respect to $e$,
up to $e_F(\epsilon)$:
\be
\Omega_{\epsilon}(y)=\int_{-\infty}^{e_F(\epsilon)} d e\,
\rho(e,y)
\label{EVdist}
\ee
In practice the integrand is obtained by interpolating
$\rho(e,y)$ from the previously stored two-dimensional grid
in the $[e,y]$ plane.
The resulting $\Omega_{\epsilon}(y)$ is shown in Fig.~3.
and compared with the corresponding WKB density,
\be
\Omega_\epsilon^{WKB} (y)=\int_{-\infty}^{e_F^{WKB}(\epsilon)} d e\,
\rho_{WKB}(e,y) = {1\over\pi} \sqrt{e_F^{WKB} - v(y)}
\label{EVdistWKB}
\ee

The contribution of the non-perturbative effects in
physical observables can most easily be seen by computing loop
averages for positive  $l$:
\be
W_{\epsilon}(l)={1\over{Z_{\epsilon}}}\int_{-\infty}^\infty
d \eta\, \Omega_{\epsilon}(\eta) e^{l \eta}
\label{Loop}
\ee
where
\be
Z_{\epsilon}=\int_{-\infty}^\infty d \eta\,
\Omega_{\epsilon}(\eta)
\label{ZLoop}
\ee
The integral in \eqr{Loop} converges fast on both ends. For large
positive
$y$, one is in the classically forbidden region,
where $\Omega_{\epsilon}$ decreases faster than exponentially
(see Fig.~3), and for large negative
$y$ the exponential factor in the integrand ensures exponential
convergence.
Thus in practice it turns out to be sufficient to take a finite upper
limit
of the integration at
$y=y_{right}$, i.e. the value of $y$ at which
$\Omega_{\epsilon}(y)$ becomes sufficiently small.

Comparison of $W_{\epsilon}(l)$ with its WKB analogue, defined by
eqs.~\eqr{Loop},\eqr{ZLoop} with $\Omega_{\epsilon}^{WKB}(y)$ instead
of $\Omega_{\epsilon}(y)$, shows that the non-perturbative part of
spectral density takes over the WKB part for large $l$.  The results
for $W_{\epsilon}(l)$ for $ 0 < l < 6 $ are shown in Fig.~4.

Note,  that the  exact  loop average,  unlike  the WKB  one,
reaches  the  minimum, and  then  grows!  This is  a  direct
consequence of the fact  that the scaling eigenvalue density
$\Omega_\epsilon(y)$   is   different  from  zero   in   the
classically forbidden region $y > 0$.

In physical terms, this  striking phenomenon is explained as
follows.  The larger the boundary $l$ of our two dimensional
space, the more area it  could encircle, and hence, the more
handles could  be attached  to it.   The number  of surfaces
with unlimited genus grows as exponential of the fifth power
of the  area \cite{GM},  and a   typical  area grows  as the
square of the length of its boundary.  So, we might expect a
very fast growth of the  loop average due to nonperturbative
contributions.

Another comment:
in the literature, doubts were raised \cite{FDavidb} , whether the
Marinari-Parisi prescription (in the WKB approximation) could
preserve
the positivity of the loop average. We did not observe such
pathologies. Moreover, we believe  that with correct nonperturbative
definition of the loop average  the positivity is guaranteed.

The point is that the spectrum of the random matrix in the
Marinari-Parisi model is infinite, it covers the whole real axis. Our
spectral integral for the loop average is manifestly positive
definite.

However, in the WKB approximation, studied in~\cite{FDavidb}, there
are several disconnected regions of the classically allowed motion
with gaps in between.  Our prescription is to {\it add } all these
contributions with positive sign. Being rewritten as a contour
integral, this would produce a set of loops encircling each allowed
region anticklockwise, or a single contour going in the imaginary
direction to the right of all the regions.

For positive $l$ one can close the contour in the left half plane,
yielding the original spectral integral. For negative $l$ one
could close the contour in the right half plane, yielding zero,
as there are no singularities of the integrand.

\bigskip
\bigskip
\begin{center}
{\LARGE\bf Acknowledgements}
\end{center}
\bigskip
\bigskip

This research was supported in part
by grant No.~90-00342 from the United States-Israel
Binational Science Foundation (BSF), Jerusalem, Israel,
and by the Basic Research Foundation administered by the
Israel Academy of Sciences and Humanities. One of us (A.M.)
would like to thank the Institute for Advanced Study at the
Tel-Aviv University for hospitality.
This work was partially supported
by the National Science Foundation under contract PHYS-90-21984.

\eject
\noindent
{\LARGE \bf Figure Captions}
\begin{itemize}
\item{Figure 1.} The particle density $\rho(e,y)$,
solution of eq.~\eqr{rhoeq}, for $\epsilon=1$ and $e=-{2\over3}$.
Dash-dotted line denotes the WKB solution.
\item{Figure 2.} The Fermi energy, $e_F(\epsilon)$, obtained from
eq.~\eqr{ImplicitEf}. The diamonds denote the actual values
computed, the continuous curve is plotted to guide the eye.
Dash-dotted line denotes the Fermi energy in the WKB approximation,
$e_F^{WKB}(\epsilon)$.
\item{Figure 3.} The scaling eigenvalue density $\Omega_\epsilon(y)$.
The WKB approximation is plotted as a dash-dotted line.
\\\\
{\bf a)}  $\Omega_{\epsilon}(y)$ and  $\Omega^{WKB}_{\epsilon}(y)$,
 $\epsilon=-1$, 0, 1.
\\\\
{\bf b)} $\Delta\Omega_{\epsilon}(y)=
\Omega^{WKB}_{\epsilon}(y) - \Omega_{\epsilon}(y)$,
 $\epsilon=-1$, 0, 1.
\\
\item{Figure 4.}
Loop average $W_\epsilon(l)$, eq.~\eqr{Loop}, for
 $\epsilon=-1$, 0, 1.
The WKB approximation is plotted as a dash-dotted line.
\end{itemize}
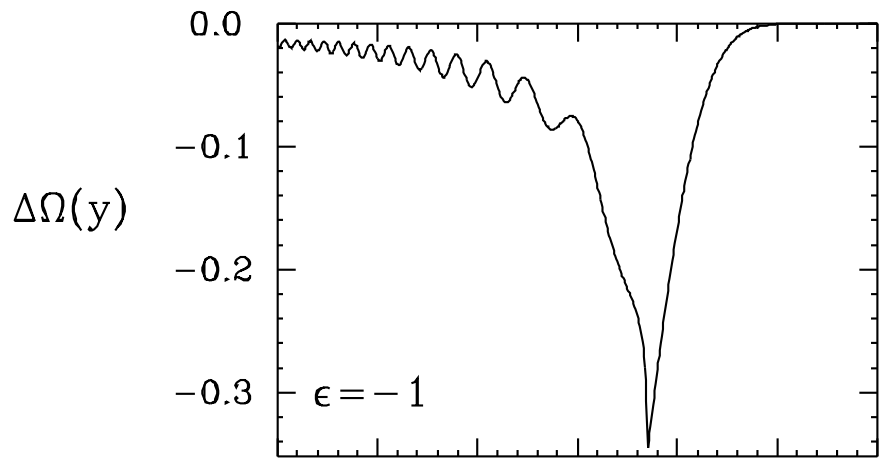
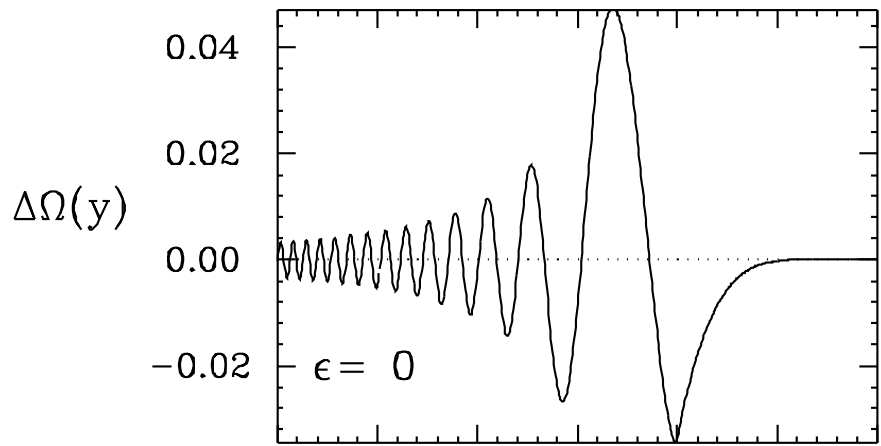
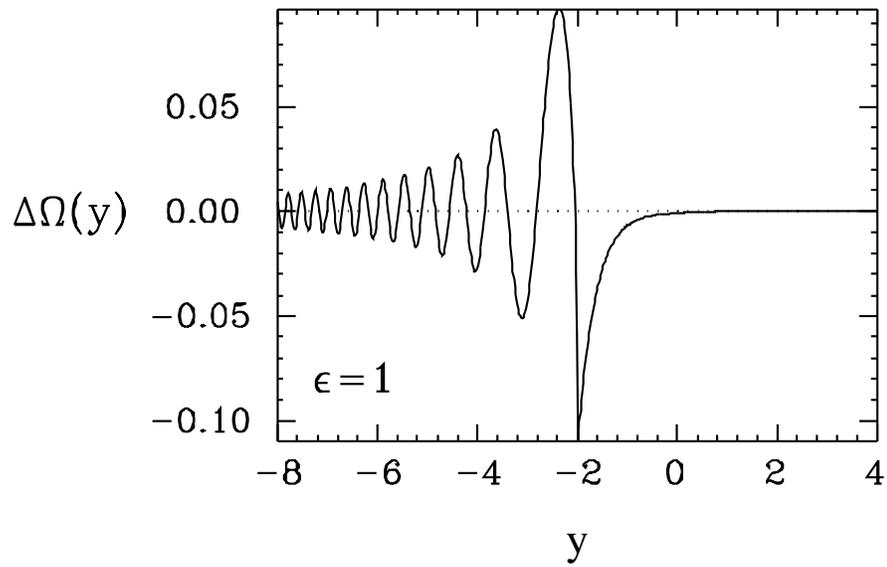
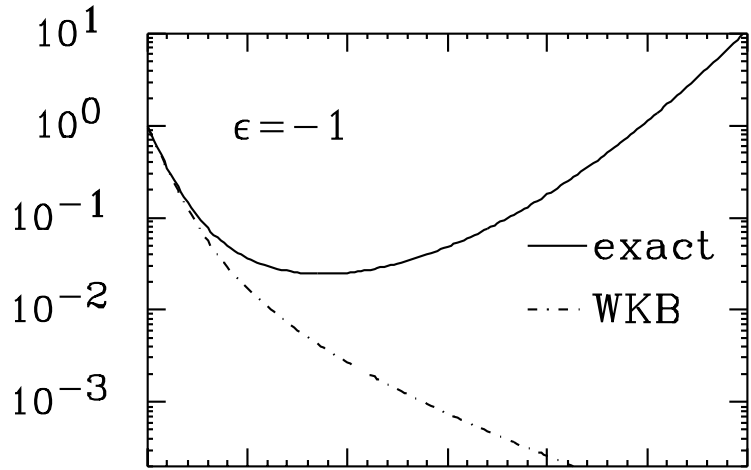
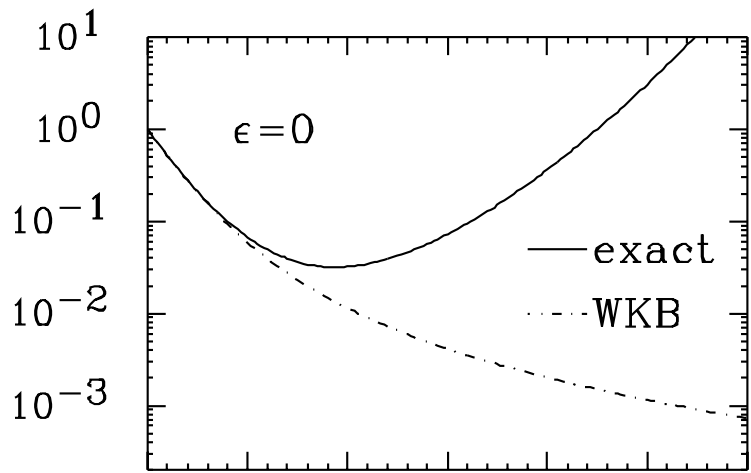
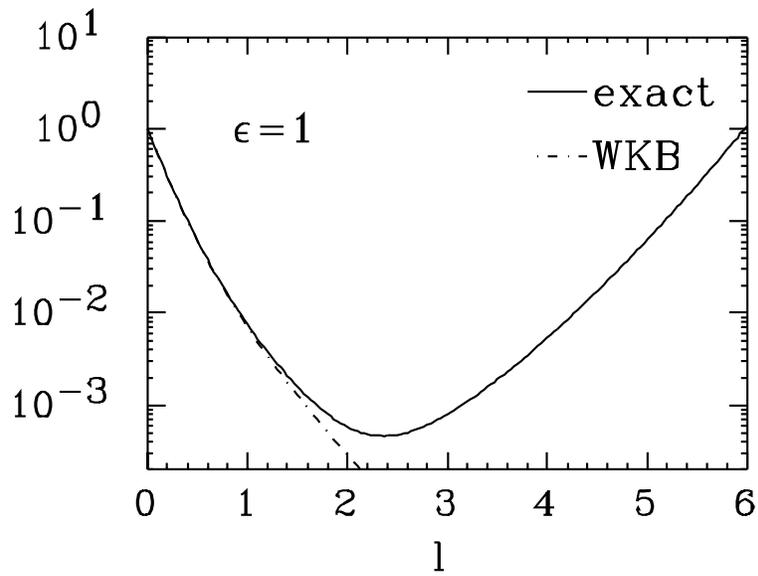
\end{document}